\documentclass{amsart}

\makeatletter
 \def\LaTeX{\leavevmode L\raise.42ex
   \hbox{\kern-.3em\size{\sf@size}{0pt}\selectfont A}\kern-.15em\TeX}
\makeatother

\newcommand{\BibTeX}{{\rm B\kern-.05em{\sc
i\kern-.025emb}\kern-.08em\TeX}}

\newtheorem{theorem}{Theorem}[section]
\newtheorem{lemma}[theorem]{Lemma}

\theoremstyle{definition}
\newtheorem{definition}{Definition}

\numberwithin{equation}{section}

\begin{document}

\title[Multiresolution Analysis on Riemannian manifolds]{Multiresolution Analysis on compact  Riemannian manifolds }

\maketitle
\begin{center}

\author{Isaac Z. Pesenson }\footnote{ Department of Mathematics, Temple University,
 Philadelphia,
PA 19122; pesenson@temple.edu. The author was supported in
part by the National Geospatial-Intelligence Agency University
Research Initiative (NURI), grant HM1582-08-1-0019. }

\end{center}

\section{Introduction}

	The problem of representation and analysis of manifold defined functions (signals, images, and data in general) is ubiquities in neuroscience, medical and biological applications. In the context of modeling the computations of the cortex, some twenty years ago, Mumford noted:  ``... \emph{the set of higher level concepts will automatically have geometric structure}". 
	Indeed, in Vision input images can be thought of as points in a high-dimensional measurement space (with each input dimension corresponding to the activity of retinal ganglion cells whose axons project from the eye to the brain), however, perceptually meaningful structures lay on a manifold embedded in this space \cite{TSL}. 	In a general context, structural and functional connectivity of the brain are usually described within network theory [Chapters XX in this volume by D. Bassett et al. \cite{BSie}; by M. Pesenson \cite{PM}; and by P. Ninez et al.  \cite{NSI}]. However, when a network of perceptual neurons can be considered as a discrete approximation of a manifold, multiresolution analysis of manifold defined functions becomes a powerful tool. 
	
	In the last decade, the importance of these and other applications triggered the development of various generalized wavelet bases suitable for the unit spheres $S^{2}$ and $S^{3}$ and  the rotation group of $\mathbb{R}^{3}$. The goal of the present study is to describe a generalization of those approaches by constructing bandlimited and localized frames in a space $L_{2}(M)$, where $M$ is a compact Riemannian manifold. The following classes of manifolds will be considered: compact manifolds without boundary, compact homogeneous manifolds, bounded domains with smooth boundaries in Euclidean spaces. 
	
	One can think of a manifold as of a surface in a Euclidean space. A homogeneous manifold is a surface with "many" symmetries like the sphere $x_{1}^{2}+...+x_{d}^{2}=1$ in Euclidean space $\mathbb{R}^{d}$. An important example of a bounded domain is a ball $x_{1}^{2}+...+x_{d}^{2}\leq 1$  in  $\mathbb{R}^{d}$. 

As it will be demonstrated below, our construction   of frames in a function space $L_{2}(M)$ heavily depends on  a proper sampling of a manifold $M$ itself. However, it is worth  stressing  that our main objective is  the $\textit{sampling  of  functions on manifolds} $.

In sections 2- 6 it is shown how to construct on  a compact manifold (with or without boundary) a "nearly"  tight bandlimited and  strongly localized frame. In other words, on a very fine scale members of our frame look almost like Dirac measures.
 In section 7 we consider the case of compact homogeneous manifolds, i.e.  which have the  form $M=G/H$, where $G$ is a compact Lie group and $H$ is its closed subgroup. For such manifolds we are able to construct tight, bandlimited and localized frames. The crucial role in this development is played by  positive cubature formulas (Theorem \ref{cubature}) and to the product property (Theorem \ref{product}), which were proved in \cite{gpes}. 
Cubature formula with positive weights in Theorem \ref{cubature} was proved 
in \cite{gpes} for general compact manifolds without boundary, but essentially the same proof can be used to establish it for manifolds with boundaries.

 In section 8 we show that properly introduced variational splines on manifolds  can be used for effective reconstruction of functions from their frame projections.

The most  important fact for our construction of frames  is that in a space of $\omega$-bandlimited functions 
the continuous and discrete norms are equivalent. This result in the case of compact and non-compact manifolds of bounded geometry was first discovered and explored in many ways in our papers \cite{Pes98}-\cite{Pes09}. In the classical cases of straight line $\mathbb{R}$ and  circle $\mathbb{S}$ the corresponding results are known as Plancherel-Polya and Marcinkiewicz-Zygmund inequalities. Our generalization of Plancherel-Polya and Marcinkiewicz-Zygmund inequalities  implies  that $\omega$-bandlimited functions  on manifolds of bounded geometry are completely determined by their vales on discrete sets of points "uniformly" distributed over $M$ with a spacing comparable to $1/\sqrt{\omega}$ and can be completely reconstructed in a stable way from their values on such sets. The last statement is obviously an extension of the famous Shannon sampling theorem to the case of Riemannian manifolds.

  Our article is a summary of some results for compact manifolds that were obtained in   \cite{gm1}-\cite{gpes}, \cite{Pes00}-\cite{Pesssubm}. To the best of our knowledge these are the pioneering papers which contain the most general   results about frames and Shannon sampling on compact and non-compact Riemannian manifolds. In particular, the paper \cite{gpes} gives an end point construction of tight localized frames on homogeneous compact manifolds.  The paper \cite{Pessubm} is the first systematic development of locali zed frames on domains in Euclidean spaces.

  Other papers which contain results about frames and wavelets on compact Riemannian manifolds  are \cite{SS},  \cite{FM1},  \cite{NPW}, \cite{PX}. In the setting of  compact manifolds necessary conditions for sampling and interpolation  in terms of Beurling-Landau densities were obtained in \cite{OP}.

Applications of frames on manifolds to scattering theory,  to statistics and cosmology can be found in  \cite{BKMP1},  \cite{BKMP2} \cite{GM100}, \cite{GM2}, \cite{gm5}, \cite{M2}-\cite{M-all}.  There is also a number of papers in which different kind of wavelets and frames are developed on non-compact homogeneous manifolds and in particular on Lie groups, see, e.g.,  \cite{CGSM},  \cite{CO1}-\cite{CKP}, \cite{FuhrM}-\cite{HF}, \cite{A}, \cite{AI}, \cite{Pes98}.

\section{ Manifolds and operators}

We describe all situations which will be discussed in the paper.

\subsection{ Manifolds without boundary.}

We will work with  second-order differential self-adjoint and non-negative  definite operators  on compact  manifolds without boundary.
The best known example of such operator is the Laplace-Beltrami which is given in a local coordinate system  by the formula
$$
L
f=\sum_{m,k}\frac{1}{\sqrt{det(g_{ij})}}\partial_{m}\left(\sqrt{det(g_{ij})}
g^{mk}\partial_{k}f\right)
$$
where $g_{ij}$ are components of the metric tensor,$\>\>det(g_{ij})$ is the determinant of the matrix $(g_{ij})$, $\>\>g^{mk}$ components of the matrix inverse to $(g_{ij})$. Spectral properties of this operator reflect (to a certain extent) metric properties of the manifold.
It is known that  Laplace-Beltrami  is a self-adjoint positive definite
operator in the corresponding space $L_{2}(M)$ constructed  from
$g$. Domains of the powers
 $L^{s/2}, s\in \mathbb{R},$ coincide with the Sobolev spaces
$H^{s}(M), s\in \mathbb{R}$. 
Since  $L$ is a second-order differential self-adjoint and non-negative  definite operator  on a compact connected  Riemannian manifold (with or without boundary), then $L$ has  a discrete spectrum $0=\lambda_{0}<\lambda_{1}\leq \lambda_{2},...$ which goes to infinity.

\subsection { Compact homogeneous manifolds}

The most complete results will be obtained for compact homogeneous manifolds.

A homogeneous compact manifold $M$ is a
$C^{\infty}$-compact manifold on which a compact
Lie group $G$ acts transitively. In this case $M$ is necessary of the form $G/K$,
where $K$ is a closed subgroup of $G$. The notation $L_{2}(M),$ is used for the usual Hilbert spaces, where $dx$ is an invariant
measure.

If $\textbf{g}$ is the Lie algebra of a compact Lie group $G$ then  it is a direct sum
$\textbf{g}=\textbf{a}+[\textbf{g},\textbf{g}]$, where
$\textbf{a}$ is the center of $\textbf{g}$, and
$[\textbf{g},\textbf{g}]$ is a semi-simple algebra. Let $Q$ be a
positive-definite quadratic form on $\textbf{g}$ which, on
$[\textbf{g},\textbf{g}]$, is opposite to the Killing form. Let
$X_{1},...,X_{d}$ be a basis of
$\textbf{g}$, which is orthonormal with respect to $Q$.
 Since the form $Q$ is $Ad(G)$-invariant, the operator
$$
-X_{1}^{2}-X_{2}^{2}-\    ... -X_{d}^{2},    \ d=dim\ G
$$
is a bi-invariant operator on $G$, which is known as the Casimir operator. This implies in particular that
the
   corresponding operator on $L_{2}(M), $
\begin{equation}\label{Casimir}
L=-D_{1}^{2}- D_{2}^{2}- ...- D_{d}^{2}, \>\>\>
       D_{j}=D_{X_{j}}, \        d=dim \ G,
\end{equation}
commutes with all operators $D_{j}=D_{X_{j}}$.
Operator $L$, which is usually called the Laplace operator, is
the image of the Casimir operator under differential of quazi-regular representation in $L_{2}(M)$.
Note that if $M=G/K$ is a compact symmetric space then the number
$d=dim\> G$ of operators in the formula (\ref{Casimir}) can be
strictly bigger than the dimension $ m=dim\> M$. For example on a
two-dimensional sphere $\mathbb{S}^{2}$ the Laplace-Beltrami
operator $L_{\mathbb{S}^{2}}$ is written as
\begin{equation}
L_{\mathbb{S}^{2}}=D_{1}^{2}+ D_{2}^{2}+
D_{3}^{2},\label{S-Laplacian}
\end{equation}
where $D_{i}, i=1,2,3,$ generates a rotation  in $\mathbb{R}^{3}$
around coordinate axis $x_{i}$:
\begin{equation}
D_{i}=x_{j}\partial_{k}-x_{k}\partial_{j},
\end{equation}
where $j,k\neq i.$

It is important to realize that in general, the operator $L$ is not necessarily the Laplace-Beltrami operator of the natural  invariant metric on $M$. But it coincides with such operator at least in the following cases:

1) If $M$ is a $d$-dimensional torus, and $(-L)$ is the sum
of squares of partial derivatives.

2) If the manifold $M$ is itself a group $G$ which is compact and
semi-simple, then $L$ is exactly the Laplace-Beltrami
operator of an invariant metric on $G$ (\cite{H2}, Ch. II).

3) If $M=G/K$ is a compact symmetric space of
  rank one, then the operator
 $L$ is proportional to the Laplace-Beltrami operator
of an invariant metric on $G/K$. This follows from the fact that, in
the rank one case, every second-order operator  which commutes with
all isometries $x\rightarrow g\cdot x, \>\>\>x\in  M,\>\>\> g\in
G,$ is proportional to the Laplace-Beltrami operator (\cite{H2},
Ch. II, Theorem 4.11).

\subsection{Bounded domains with smooth boundaries}
 
 Our consideration includes also  an open  bounded domain $M\subset \mathbb{R}^{d}$ with a smooth boundary $\Gamma$ which is a smooth $(d-1)$-dimensional oriented manifold.  Let $\overline{M}=M\cup \Gamma$ and $L_{2}(M)$ be  the  space of functions  square-integrable with respect to the Lebesgue  measure $dx=dx_{1}...dx_{d}$ with the norm denoted as $\|\cdot \|$. If $k$ is a natural number the notations $H^{k}(M)$ will be  used for  the Sobolev space of distributions on $M$ with the norm
 $$
 \|f\|_{H^{k
 }(M)}=\left(\|f\|^{2}+\sum _{1\leq |\alpha |\leq k}\|\partial^{|\alpha|} f\|^{2}\right)^{1/2}
 $$
where $\alpha=(\alpha_{1},...,\alpha_{d})$ and $\partial^{|\alpha|}$ is a mixed partial derivative 
$$
\left(\frac{\partial}{\partial x_{1}}\right)^{\alpha_{1}}...\left(\frac{\partial}{\partial x_{d}}\right)^{\alpha_{d}}.
$$
Under our assumptions the space $C^{\infty}_{0}(\overline{M})$ of infinitely smooth functions with support in $\overline{M}$  is dense  
in $H^{k}(M)$. Closure in $H^{k}(M)$ of the space   $C_{0}^{\infty}(M)$ of smooth functions with support in $M$ will be denoted as $H_{0}^{k}(M)$.

 Since $\Gamma$ can be treated as a smooth Riemannian manifold one can introduce Sobolev scale of spaces $H^{s}(\Gamma),\>\> s\in \mathbb{R},$ as, for example, the domains of the Laplace-Beltrami operator $L$ of a Riemannian metric on $\Gamma$.
 
 According to the trace theorem there exists a well defined continuous surjective trace operator 
 $$
 \gamma: H^{s}(M)\rightarrow H^{s-1/2}(\Gamma),\>\>s>1/2,
 $$
 such that for all functions $f$  in $H^{s}(M)$ which are smooth up to the boundary the value $\gamma f$ is simply a restriction of $f$ to $\Gamma$.

 One considers the following  operator 
 \begin{equation}\label{Op}
 Pf=-\sum_{j,k}\partial_{j}\left(a_{j,k}(x)\partial_{k}f\right),
\end{equation}
with coefficients in $C^{\infty}(M)$ where the matrix $(a_{j,k}(x))$ is real, symmetric  and positive definite on $\overline{M}$.
 The operator $L$ is defined as the Friedrichs extension of $P$, initially defined on $C_{0}^{\infty}(M)$, to the set of all functions $f$ in $H^{2}(M)$ with constraint $\gamma f=0$. The Green formula implies that this operator is self-adjoint. It is also a positive operator and the domain of its positive square root $L^{1/2}$ is the set of all functions $f$ in $H^{1}(M)$ for which $\gamma f=0$. 
  
 Thus, one has a self-adjoint positive definite operator in the Hilbert space $L_{2}(M)$ with a discrete spectrum $0<\lambda_{1}\leq \lambda_{2},...$ which goes to infinity.  
\bigskip

\section{Hilbert frames}

If a manifold $M$ is compact and one has an  
elliptic, self-adjoint positive definite operator $L$ in $L_{2}(M)$   then $L$  has a discrete spectrum $0=\lambda_{0}<\lambda_{1}\leq \lambda_{2}\leq......$ which goes to infinity and there exists a family  $\{u_{j}\}$  of orthonormal eigenfunctions which form a  basis in $L_{2}(M)$.  
Since eigenfunctions have perfect localization properties in the spectral domain they cannot be localized on the manifold. 

It is the goal of our article to construct "better bases" in corresponding $L_{2}(M)$ spaces which will have rather strong localization on a manifold and  in the spectral domain.

In fact, the "kind of basis" which we are going to construct   is known today as a frame.

A set of vectors $\{\theta_{v}\}$  in a Hilbert space $H$ is called a frame if there exist constants $A, B>0$ such that for all $f\in H$ 
\begin{equation}
A\|f\|^{2}_{2}\leq \sum_{v}\left|\left<f,\theta_{v}\right>\right|^{2}     \leq B\|f\|_{2}^{2}.
\end{equation}
The largest $A$ and smallest $B$ are called lower and upper frame bounds.

The set of scalars $\{\left<f,\theta_{v}\right>\}$ represents a set of measurements of a signal $f$. To synthesize signal $f$ from this set of measurements one has to find another (dual) frame $\{\Theta_{v}\}$ and then a reconstruction formula is 
\begin{equation}
f=\sum_{v}\left<f,\theta_{v}\right>\Theta_{v}.
\end{equation}

Dual frame is not unique in general. Moreover it is difficult to find a dual frame.

If in particular $A=B=1$ the frame is said to be  tight   or Parseval. 
Parseval frames are similar in many respects to orthonormal wavelet bases.  For example, if in addition all vectors $\theta_{v}$ are unit vectors, then the frame is an  orthonormal basis.

The main feature of Parseval frames is that 
decomposing
and synthesizing a signal or image from known data are tasks carried out with
the same set of functions. 
The important differences between frames and, say, orthonormal bases is their  redundancy that helps reduce the effect of noise in data.

Frames in Hilbert spaces of functions  whose members have simultaneous localization in space and frequency  arise naturally in wavelet analysis on Euclidean spaces  when continuous wavelet transforms are discretized.
Such  frames have been constructed, studied, and
employed extensively in both theoretical and applied problems (in addition to papers listed in introduction one can refer to  \cite{D}, \cite{FY}, \cite{M1},  \cite{M2}).

\section{Multiresolution and sampling}

\bigskip

On a very general level Multiresolution Analysis on Riemannian manifolds can be described as a framework that brings together metric properties (geometry) of a manifold and spectral properties (Fourier analysis)  of the corresponding Laplace-Beltrami operator.

The objective of our work is to construct a  frame $\Psi_{l}=\{\psi_{l,j}\}$  in the space  $L_{2}(M)$  which  is somewhere  between   the two "extreme" bases i.e. eigenfunctions of the Laplace -Beltrami operator  and a collection of Dirac functions. 
Multiresolution Analysis  is using   discretization of a manifold as a way to link geometry and analysis.
To be more precise, one approximates a manifold (space) by 
 sets of points and associates with such a discretization a frame in the space $L_{2}(M)$ which is localized in frequency and on manifold.

 \begin{definition}\label{band}
In the case of compact manifolds (with or without boundary) 
for a given $\omega>0$, the span of eigenfunctions $u_{j}$ 
 $$
 Lu_{j}=\lambda_{j}u_{j}
 $$
 with $\lambda_{j}\leq \omega$ is denoted as  $E_{\omega}(L)$ and is called the space of bandlimited functions on $M$ of bandwidth $\omega$.
 \end{definition}
According to the  Weyl's asymptotic formula \cite{Hor} one has
\begin{equation}\label{Weyl}
dim \>E_{\omega}(L)\sim C\> Vol(M)\omega^{d/2},
\end{equation}
where $d=dim \>M$ and $C$ is an absolute constant.

As it was already mentioned,  the important fact is that $\omega$-bandlimited functions  are completely determined by their vales on discrete sets of points "uniformly" distributed over $M$ with a spacing comparable to $1/\sqrt{\omega}$ and can be completely reconstructed in a stable way from their values on such sets.

Intuitively, such discrete sets can be associated with a scale $1/\sqrt{\omega}$: a finer scaling requires larger frequencies.

 The main objective of Multiresolution Analysis is to construct  a  frame in $L_{2}(M)$ whose structure reflects the   relation between scaling and frequency.

 Now  we introduce  what can be considered as a notion of  "points uniformly distributed over a manifold".

 One can show that for a Riemannian manifold  $M$  of bounded geometry there exists
a natural number $N_{M}$ such that  for any sufficiently small
$\rho>0$ there exists a set of points $\{y_{\nu}\}$ such that:

1)  the balls $B(y_{\nu}, \rho/4)$ are disjoint,

2) the balls $B(y_{\nu}, \rho/2)$ form a cover of $M$,

3) the multiplicity of the cover by balls $B(y_{\nu},\rho)$
is not greater $N_{M}.$

\begin{definition}
A set of points $M_{\rho}=\{y_{\nu}\}$ is called a $\rho$-lattice if it is a set of centers of balls with the above listed properties 1)- 3).
\end{definition}

\bigskip

Our main result can be described as  follows.

Given a Riemannian manifold $M$ and a sequence of positive numbers $\omega_{j}=2^{2j+1},\>\>j=0,1,....$, we  consider the Paley-Wiener space $E_{\omega_{j}}(L)$ of functions bandlimited to $[0, \omega_{j}]$ and  for  a specific $c_{0}=c_{0}(M)$ consider  a set of scales 
$$
\rho_{j}=c_{0}\omega_{j}^{-1/2},\>\>\omega_{j}=2^{2j+1},\>\>j=0,1,....,
$$ 
and construct a corresponding set of lattices 
$$
M_{\rho_{j}}=\{x_{j,\>k}\}_{k=1}^{m_{j}}, \>\>x_{j,\>k} \in M,\>\>k\in \mathbb{Z},\>\>dist\>(x_{j, k_{1}},\>\>x_{j,k_{2}})\sim \rho_{j},
$$
of points which are  distributed over $M$ with a spacing  comparable to $\rho_{j}$. 

\bigskip

With every point $x_{j,\>k}$ we associate a function  $\Theta_{j,\>k}\in L_{2}(M)$  such that

\begin{enumerate}

\item 

 function  $\Theta_{j,k},\>\>1\leq k\leq m_{j}\in \mathbb{N},$  is bandlimited to $[0, \>\>\omega_{j}]$;

  \item
  
  the "essential" support of $\Theta_{j,k},\>\>1\leq j\leq m_{k}\in \mathbb{N},$  is in the ball $B(x_{j,k},\rho_{j})$  with center at $x_{j,\>k}$ and of radius $\rho_{j}$;

\item 

 the set  $\bigcup_{k=1}^{m_{j}}\Theta _{j,k}$ is a frame in  $E_{\omega_{j}}(L)$;

 \item the set  $\bigcup_{j=0}^{\infty}\bigcup_{k=1}^{m_{j}}\Theta_{j,k}$ is a frame   in $L_{2}(M)$;

\end{enumerate}

Note that $\Theta=\bigcup_{j,k}\{\Theta_{j,k}\}$ corresponds to the set $X=\bigcup_{j}M_{\rho_{j}}$ which is union of all scales.

Thus, by changing a subset of functions $\Theta_{j}=\bigcup_{k=1}^{m_{j}}\Theta_{j,k}$  to a subset  $\Theta_{i}=\bigcup_{k=1}^{m_{i}}\Theta_{i,k}$ (in the space $L_{2}(M)$) we  are actually going

\bigskip

a)  from the scale $M_{\rho_{j}}$ to the scale $M_{\rho_{i}}$ in space 

\bigskip

and at the same time

\bigskip

b)  from the  frequency band $[0, \>\>2^{2j+1}]$ to the  frequency band $[0, \>\>2^{2i+1}]$  in the frequency domain.

Intuitively,  index $1\leq k\leq m_{j}$ corresponds to  Dirac measures (points on a manifold) and $j\in  \mathbb{Z}$ to bands of frequencies.

\section{Shannon Sampling of bandlimited functions on Manifolds }

The most important fact for our development is an analog of the Shannon's Sampling Theorem for Riemannian manifolds of bounded geometry, which was established in our papers  \cite{Pes98}-\cite{Pes09}.  Our generalization of the Sampling Theorem states  that $\omega$-bandlimited functions on a manifold $M$  are completely determined by their values on  sets of points  distributed over $M$ with a spacing comparable to $1/\sqrt{\omega}$ and can be completely reconstructed in a stable way from their values on such sets.

\begin{theorem}
\label{sampling}

 For  a given  $0<\delta<1$ there exists  a constant $c_{0}=c_{0}(M,\>\delta)$ such that, if
 \begin{equation}\label{samrate}
 \rho=c_{0} \omega^{-1/2},\>\>\>\omega>0,
 \end{equation}
 then for any $\rho$-lattice $M_{\rho}=\{x_{k}\}$ there exists a set of weights $\mu_{k}(\rho)\sim \rho^{d}$ such that one has the following Plancherel-Polya inequalities (or frame inequalities) 
\begin{equation}\label{sam}
(1-\delta) \|f\|^{2}\leq 
\sum _{k}\mu_{k}(\rho)|f(x_{k})|^{2}\leq \|f\|^{2}.
\end{equation}
for all $f\in E_{\omega}( L)$.

 The inequalities (\ref{sam}) imply that every $f\in E_{\omega}( L)$ is uniquely determined by its values on $M_{\rho}=\{x_{k}\}$ and can be reconstructed from these values in a stable way.

\end{theorem}

It shows that if  $\theta_{k}$ is the orthogonal projection of the distribution $\sqrt{\mu_{k}(\rho)}\delta_{x_{k}}$ ($\delta_{x_{k}}$ is the Dirac measure  at $x_{k}$) on the space $E_{\omega}(L)$,  then the following frame
inequalities hold
$$
(1-\delta) \|f\|^{2}\leq 
\sum _{k}|\left<f,\>\theta_{k}\right>|^{2}\leq \|f\|^{2}
$$
for all $f\in E_{\omega}( L)$.
In other words, we obtain that the set of functions $\{\theta_{k}\}$ is a frame in the space $E_{\omega}(L)$.

According to the general theory of frames,  one has that if $\{\Psi_{k}\}$ is a frame which is dual to $\{\theta_{k}\}$ in the space $E_{\omega}(L)$ (such frame is not unique) then the following reconstruction formula holds
\begin{equation}\label{recon}
f=\sum_{k}\left<f,\theta_{k}\right>\Psi_{k}.
\end{equation}

The condition  (\ref{samrate}) imposes a specific rate of sampling in (\ref{sam}).
It is interesting to note 
 that this rate is essentially optimal. Indeed, on one hand   the  Weyl's asymptotic formula  (\ref{Weyl}) gives the dimension of the space $E_{\omega}(L)$. 
On the other hand,  the condition  (\ref{samrate}) and the definition of a $\rho$-lattice imply  that 
the number of points in an "optimal" lattice $M_{\rho}$  for (\ref{sam}) can be  approximately estimated as
$$
 card\>M_{\rho}\sim\frac{Vol(M)}{c_{0}^{d}\omega^{-d/2}}=c\>Vol(M)\omega^{d/2},\>\>\>d=dim\>M,
$$
which is in agreement with  the Weyl's formula.

\section{Localized frames on compact manifolds}

 In this section for  every $f\in L_{2}(M)$  we construct a special decomposition into bandlimited functions and then perform a discretization step by applying the Theorem \ref{sampling} from the previous section.

 Choose a function $\Phi \in C_c^{\infty}(\mathbb{R}^{+})$, supported in the interval $[2^{-2},2^4]$ such that
\begin{equation}
\label{addto1}
\sum_{j=-\infty}^{\infty} |\Phi(2^{-2j}s)|^2 = 1
\end{equation}
for all $s > 0$.  

For example, we could choose a smooth monotonically decreasing function 
   $\psi$ on $\mathbb{R}^{+}$ with $0 \leq \psi \leq 1$, with $\psi \equiv 1$
in $[0,2^{-2}]$ and with $\psi = 0$ in $[2^2,\infty)$. In this case  $\psi(s/2^2) - \psi(s)\geq 0$ and we 
set 
\begin{equation}\label{function}
\Phi(s) = [\psi(s/2^2) - \psi(s)]^{1/2},\>\>\>s > 0,
\end{equation}
which will have support in $[2^{-2},\> 2^{4}]$ and will satisfy (\ref{addto1}).
  Using the Spectral Theorem for $L$ and the equality (\ref{addto1}) one can obtain 
\begin{equation}
\label{addto1op}
\sum_{j=-\infty}^{\infty} |\Phi|^2({2^{-2j}L}) = I-P, 
\end{equation}
where $P$ is the projector on the kernel of $L$ and 
where the sum (of operators) converges strongly on $L_2(M)$.

 It should be noted, that  in  the case of Dirichlet boundary conditions $P=0$.
 
 Formula (\ref{addto1op}) implies the following equality

\begin{equation}
\label{norm equality}
\sum_{j\in \mathbb{Z}} \|\Phi({2^{-2j} L})f\|^2_2 = \|(I-P)f\|^2_2.
\end{equation}
Moreover, since function $ \Phi(2^{-2j} s)$ has support in  $
[2^{2j-2},\>\>2^{2j+4}]$ the function $\Phi({2^{-2j} L})f $ is bandlimited to  $
[2^{2j-2},\>\>2^{2j+4}]$. 

According to Theorem \ref{sampling} for a fixed $0<\delta<1$ there exists a constant $c_{0}(M,\>\delta)$ such that  for  
$$
\rho_{j}=
c_{0}\omega_{j}^{-1/2}=c_{0}2^{-j-2},\>\>\>j\in \mathbb{Z},
$$
and  any $\rho_{j}$-lattice $M_{\rho_{j}}=\{x_{j,k}\},\>\>1\leq k\leq K_{j},$ the inequalities (\ref{sam}) hold.

Thus, if  $\theta_{j,k}\in E_{\omega_{j}}(L)=E_{2^{2j+4}}(L)$ is the projection of $\sqrt{\mu_{k}(\rho)}\delta_{x_{k}}$onto $E_{\omega_{j}}(L)$, then we have the following frame inequalities   in  $PW_{\omega_{j}}(M)$ for every $j\in \mathbb{Z}$
\begin{equation}
(1-\delta)\left \|\> \Phi\left({2^{-2j} L}\right)f\right\|^{2}\leq
$$
$$ 
\sum _{ k=1}^{K_{j}}\left|\left< \Phi\left({2^{-2j} L}\right)f, \theta_{j,k}\right>\right|^{2}\leq \left\| \>\Phi\left({2^{-2j} L}\right)f\right\|^{2},
\end{equation}
where  $\Phi({2^{-2j}L})f\in E_{\omega_{j}}(L)=E_{2^{2j+4}}(L)$. Together with (\ref{norm equality}) it gives for any $f\in L_{2}(M)$ the following inequalities
\begin{equation}
(1-\delta)\|f\|^{2}\leq
$$
$$
\sum_{j\in \mathbb{Z}}\>\>\sum _{ k=1}^{K_{j}}\left|\left<\Phi({2^{-2j}L})f, \theta_{j,k}\right>\right|^{2}\leq \|f\|^{2},\>\>\>f\in L_{2}(M),\>\>\>\theta_{j,k}\in E_{2^{2j+4}}(L).
\end{equation}
But, since operator $\Phi\left({2^{-2j} L}\right)$ is self-adjoint,
$$
\left<\Phi\left({2^{-2j} L}\right)f, \theta_{j,k}\right>=\left<f, \Phi\left({2^{-2j}L}\right)\theta_{j,k}\right>,
$$
we obtain, that for the functions
\begin{equation}\label{frame-functions}
\Theta_{j,k}=\Phi\left({2^{-2j} L}\right)\theta_{j,k}\in  E_{[2^{2j-2},\>\>2^{2j+4}]}(L)\subset E_{2^{2j+4}}(L)
\end{equation}
the following frame inequalities hold
\begin{equation}\label{frame-ineq}
(1-\delta)\|(I-P)f\|^{2}\leq\sum_{j\in \mathbb{Z}}\>\>\sum _{k=1}^{K_{j}}\left|\left<f, \Theta_{j,k}\right>\right|^{2}\leq \|(I-P)f\|^{2},\>\>\>\>f\in L_{2}(M).
\end{equation}

 The  next goal  is to find an explicit formula for the operator $\Phi({2^{-2j} L})$ and to 
show  localization  of the frame elements $
\Theta_{j,k}$.

According to the Spectral Theorem if a  self-adjoint positive-definite operator $L$ has a discrete spectrum $0<\lambda_{1}\leq \lambda_{2}\leq..., $ and a corresponding set of eigenfunctions $u_{j}$,
 
 $$
 Lu_{j}=\lambda_{j}u_{j},
 $$
 which form an orthonormal basis in  $L_{2}(M)$, then for any
  bounded  real-valued function $F$ of one variable  one can construct a self-adjoint operator $F(L)$  in $L_{2}(M)$ as
 \begin{equation}\label{function-1}
 F(L)f(x)=\int_{M}\mathcal{K}^{F}(x,y)f(y)dy,\>\>f\in L_{2}(M),
 \end{equation} 
 where  $\mathcal{K}^{F}(x,y)$ is a smooth function defined by the formula
 \begin{equation}\label{kernel-1}
\mathcal{K}^{F}(x,y)= \sum _{m}F(\lambda_{m})u_{m}(x)\overline{u_{m}}(y).
\end{equation}
The following notations will be   used
\begin{equation}\label{operator}
\left[F(tL)f\right](x)=\int_{M}\mathcal{K}^{F}_{\sqrt{t}}(x,y)f(y)dy,\>\>f\in L_{2}(M),
\end{equation}
 where 
 \begin{equation}\label{kernel-2}
 \mathcal{K}^{F}_{\sqrt{t}}(x,y)=\sum _{m}F(t\lambda_{m})u_{m}(x)\overline{u_{m}}(y).
 \end{equation}
 In our situation  we have  the formulas
 \begin{equation}\label{kernel-10}
 \mathcal{K}_{2^{-j}}^{\Phi}(x, y)= \sum_{m\in \mathbb{Z}_{+}} \Phi(2^{-2j}\lambda_m)u_{m}(x) \overline{u}_{m}(y),
 \end{equation}
and
\begin{equation}
\left[\Phi(2^{-2j}L)f\right](x)=\int_{M}\mathcal{K}^{\Phi}_{2^{-j}}(x,y)f(y)dy,\>\>f\in L_{2}(M).
\end{equation}

By expanding $f \in L_{2}(M)$ in terms of eigenfunctions of $L$
$$
f= \sum_{m\in \mathbb{Z}_{+}} c_m(f) u_m,\>\>\>c_m(f)=\left<f,u_{m}\right>,
$$ 
one has 
$$
\Phi({2^{-2j} L})f= \sum_{2^{2j-2}\leq \lambda_{m}\leq 2^{2j+4}} \Phi(2^{-2j}\lambda_m)c_m(f) u_m.
$$

After all we obtain
\begin{equation}
\Theta_{j,k}=\Phi\left({2^{-2j} L}\right)\theta_{j,k}=
$$
$$
\mathcal{K}^{\Phi}_{2^{-j}}(x,x_{j,k})=\sum_{m\in\mathbb{Z}_{+}} \Phi(2^{-2j}\lambda_m)c_m(\theta_{j,k})\overline{u}_{m}(x_{j,k}) u_m(x).
\end{equation}

Localization properties of the kernel $\mathcal{K}_{t}^{F}(x,y)$ are given in the following statement.

\begin{lemma}\label{localiz}

If $L$ is an elliptic self-adjoint second order differential operators on a compact manifold (without boundary or with a smooth boundary) and $\mathcal{K}_{t}^{F}(x,y)$ is given by (\ref{kernel-1}), where $F$ is an even function in $C^{\infty}_{c}(\mathbb{R})$, then  on $\Omega\times\Omega\setminus \Delta,\>\>\>\Delta=\{(x,x)\},\>x\in \Omega$, kernel $\mathcal{K}_{t}^{F}(x,y)$ vanishes to infinite order as  $t$ goes to zero.

\end{lemma}

Different proofs can be found in  \cite{gm1}, \cite{gpes},   \cite{Tay81}, \cite{Hor}.

\bigskip

The last property shows  that kernel $\mathcal{K}_{t}^{F}(x,y)$ is localized as long as $F$ is an even Schwartz function.  Indeed, if for a fixed point $x\in M$ a point $y\in M$ is "far" from $x$ and $t$ is small, then the value of $\mathcal{K}_{t}^{F}(x,y)$ is small. 

The Lemma \ref{localiz} is an analog of the important fact for Euclideant spaces that the Fourier transform of a Schwartz function is a Schwartz function. 

Since $\mathcal{K}_{t}^{\Phi}(x,y)$ is smooth and  $M$ is bounded we can  express localization of $\mathcal{K}_{t}^{\Phi}(x,y)$ by using the following inequality: 
for any $N>0$ there exists a $C(N)$ such, that for all sufficiently small positive $t$ 
  \begin{equation}
    \left|\mathcal{K}_{t}^{\Phi}(x,y)\right| \leq 
    C( N)\frac{t^{-d}}{\max(1, \>\>t^{-1}|x-y|)^{N}},\>\>\>t>0.
 \end{equation}

  From here one obtains the next inequality

 $$
\left| \Theta_{j,k}(x)\right|=\left|\Phi(2^{-2j}L)\theta_{j,k}(x)\right| =\left|\mathcal{K}^{\Phi}_{2^{-j}}(x,x_{j,k})\right|\leq 
$$
$$
  C(N) \frac{2^{dj}}{\max(1, \>\>2^{j}|x-x_{j,k}|)^{N}}.
  $$

 Thus, the following statement about localization of every $ \Theta_{j,k}$ holds.
 
  \begin{lemma}\label{fr-local}
  For  any $N>0$ there exists a $C( N)>0$ such  that 
  \begin{equation}
    |\Theta_{j,k}(x)|\leq 
    C( N) \frac{2^{dj}}{\max(1, \>\>2^{j}|x-x_{j,k}|)^{N}},
 \end{equation}
  for all $ j\in \mathbb{Z}.$ 
    \end{lemma}

    Inequality (\ref{frame-ineq}) and Lemma \ref{fr-local} give the following Frame Theorem.

\begin{theorem}\label{FrTh}
 For  a given  $0<\delta<1$ there exists a constant a constant $c_{0}=c_{0}(M,\>\delta)$ such that, if
 $$
 \rho_{j}=c_{0}2^{-j-2},\>\>\>\omega>0,\>\>j\in \mathbb{Z},
$$
and $M_{\rho_{j}}=\{x_{j,k}\},$  is a  $\rho_{j}$-lattice, then  the corresponding set of functions  $\left\{\Theta_{j,k}\right\}$:
$$
\Theta_{j,k}=\Phi\left({2^{-2j} L}\right)\theta_{j,k}, \>\>\>j\in \mathbb{Z},\>\>1\leq k\leq K_{j}, 
$$
where $\theta_{j,k}$ is  projection of the measure $\sqrt{\mu_{j,k}(\rho_{j})}\delta_{x_{j,k}}$ onto $E_{2^{2j+4}}( L)$, 
  is a frame in $L_{2}(M)$ with constants $1-\delta$ and $1$. 
  
  In other words, the following frame inequalities hold 
 $$
 (1-\delta)\|f\|^{2}\leq \sum_{j\in \mathbb{Z}}\>\> \sum_{1\leq k\leq K_{j}} \left|\left<f,\Theta_{j,k}\right>\right|^{2}\leq \|f\|^{2},
 $$
for all $f\in L_{2}(M)$.

 Every $ \Theta_{j,k}$ is bandlimited to  $[2^{2j-2}, 2^{2j+4}]$ and in particular belongs to $E_{2^{2j+4}}( L)$.  Localization properties of $ \Theta_{j,k}$ are described in Lemma \ref{fr-local}.
\end{theorem}

\section{Parseval frames on homogeneous manifolds}

In this section  we assume that a manifold $M$ is homogeneous (has many symmetries) in the sense that it is of the form  $M=G/H,$  where $G$ is a compact Lie group and $H$ is its closed subgroup.  In this situation we construct spaces of bandlimited functions by using  the Casimir operator $L$ that was defined in (\ref{Casimir}).

Under these assumptions we are able to construct a  tight bandlimited and localized frame in the space $L_{2}(M)$.

\begin{theorem}\label{product}(Product property \cite{gpes})
\label{prodthm}
If $M=G/H$ is a compact homogeneous manifold and $L$
 is the same as above, then for any $f$ and $g$ belonging
to $E_{\omega}(L)$,  their product $fg$ belongs to
$E_{4m\omega}(L)$, where $m$ is the dimension of the
group $G$.

\end{theorem}
\begin{theorem}\label{cubature} (Cubature  formula \cite{gpes})
There exists  a  positive constant $a_{0}$,    such  that if  $\rho=a_{0}(\omega+1)^{-1/2}$, then
for any $\rho$-lattice $M_{\rho}$, there exist strictly positive coefficients $\alpha_{x_{k}}>0, 
 \  x_{k}\in M_{\rho}$, \  for which the following equality holds for all functions in $ E_{\omega}(M)$:
\begin{equation}
\label{cubway}
\int_{M}fdx=\sum_{x_{k}\in M_{\rho}}\alpha_{x_{k}}f(x_{k}).
\end{equation}
Moreover, there exists constants  $\  c_{1}, \  c_{2}, $  such that  the following inequalities hold:
\begin{equation}
c_{1}\rho^{d}\leq \alpha_{x_{k}}\leq c_{2}\rho^{d}, \ d=dim\ M.
\end{equation}
\end{theorem}

Using the same notations as in the previous section we find

\begin{equation}
\label{addto1sc}
\sum_{j=-\infty}^{\infty} \|\Phi({2^{-2j} L})f\|^2_2 = \|(I-P)f\|^2_2
\end{equation}

Expanding $f \in L_{2}(M)$ in terms of eigenfunctions of $L$
$$
f= \sum_i c_i (f) u_i,\>\>\>c_i(f)=\left<f, u_{i}\right>,
$$ 
we have
$$
\Phi({2^{-2j} L})f= \sum_i \Phi(2^{-2j}\lambda_i)c_i(f) u_i.
$$

Since for every $j$ function $\Phi(2^{-2j}s)$ is supported in the interval $[2^{2j+2}, 2^{2j+4}]$ the function $\Phi({2^{-2j}L})f$ is bandlimited and belongs to $E_{2^{2j+4}}({L})$.

But then the function 
$\overline{\Phi({2^{-2j} L})f}$ is also in $E_{2^{2j+4}}({ L})$.
Since 
$$
|\Phi({2^{-2j}L})f|^2=\left[\Phi({2^{-2j} L})f\right]\left[\overline{\Phi({2^{-2j} L})f}\right],
$$
one can use  the product property to conclude that  
$$
|\Phi({2^{-2j} L})f|^2\in 
E_{4m2^{2j+4}}({ L}),
$$
where $m=dim\>G,\>\>{M}=G/H$.

To summarize, we proved, that for every $f\in L_{2}(M)$ we have the following decomposition 
\begin{equation}
\label{addto1sc}
\sum_{j=-\infty}^{\infty} \|\Phi({2^{-2j} L})f\|^2_2 = \|(I-P)f\|^2_2,\>\>\>\>\>
|\Phi({2^{-2j}L})f|^2\in 
E_{4m2^{2j+4}}({ L}).
\end{equation}

The next objective is to perform a discretization step. According to our result about cubature formula there exists a constant $a_{0}>0$ such that for all integer $j$ if  

\begin{equation}
\label{rhoj}
\rho_j = a_{0}(4m2^{2j+4}+1)^{-1/2}\sim 2^{-j},\>\>m=dim\>G, \>\>M=G/H, 
\end{equation}
then for any  $\rho_{j}$-lattice $M_{\rho_{j}}$ one can find coefficients $b_{j,k}$ with
\begin{equation}
\label{wtest1}
b_{j,k}\sim \rho_j^{d},\>\>\>d=dim\>M,
\end{equation}
for which the following exact cubature formula holds
\begin{equation}
\label{cubl2}
\|\Phi({2^{-2j} L})f\|^2_2 = \sum_{k=1}^{J_j}b_{j,k}|[\Phi({2^{-2j}L})f](x_{j,k})|^2,
\end{equation}
where $x_{j,k} \in M_{\rho_j}$, ($k = 1,\ldots,J_j = card\>(M_{\rho_j})$).

Now, for $t > 0$, let $\mathcal{K}_t^{\Phi}$ be the kernel of $\Phi(t^2{ L})$, so that,
for $f \in L_2(M)$, 
\begin{equation}
\label{kerfrm}
[\Phi(t^2{ L})] f(x) = \int_{M} \mathcal{K}^{\Phi}_t(x,y)f(y) dy.
\end{equation}
For $x,y \in M$, we have
\begin{equation}
\label{ktexp}
\mathcal{K}^{\Phi}_t(x,y) = \sum_i \Phi(t^2\lambda_i) u_i(x) \overline{u}_i(y).
\end{equation}

Corresponding to each $x_{j,k}$ we now define the functions
\begin{equation}
\label{vphijkdf}
\psi_{j,k}(y) = \overline{\mathcal{K}^{\Phi}_{2^{-j}}}(x_{j,k},y) = \sum_i \overline{\Phi}(2^{-2j}\lambda_i) \overline{u}_i(x_{j,k}) u_i(y),
\end{equation}
\begin{equation}
\label{phijkdf}
\Psi_{j,k} = \sqrt{b_{j,k}} \psi_{j,k}.
\end{equation}

We find that for all $f \in 
L_2(M)$,
\begin{equation}
\label{parfrm}
\|(I-P)f\|^2_2 = \sum_{j,k} |\langle f,\Psi_{j,k} \rangle|^2.
\end{equation}
Note that, by (\ref{vphijkdf}), (\ref{phijkdf}), and the fact that $\Phi(0) = 0$, each
$\Psi_{j,k} \in (I-P)L_2(M)$.

Thus the following statement is proved.
\begin{theorem}
If $M$ is a homogeneous manifold, then the set of functions $\{\Psi_{j,k}$\}, constructed in (\ref{phijkdf}) is a Parseval frame in the space $(I-P)L_{2}(M)$. 
\end{theorem}

Here,  functions $\Psi_{j,k}$  belong to $E_{\omega_{j}}(L)$ and their spatial decay follows from Lemma \ref{localiz}.

By general frame theory, if $f \in L_2(M)$, we have
\begin{equation}
\label{recon}
(I-P)f = \sum_{j=\Omega}^{\infty}\sum_k \langle f,\Psi_{j,k} \rangle \Psi_{j,k} = 
\sum_{j=\Omega}^{\infty}\sum_k b_{j,k} \langle f,\psi_{j,k} \rangle \psi_{j,k},
\end{equation}
with convergence in $L_2(M)$.

\section{Variational splines on manifolds}

As it was explained (see the  formula (\ref{recon}))  one can always  use a dual frame for reconstruction of a function from its projections. However, in general  it is not easy to construct a dual frame (unless the frame is tight and then it is self-dual).

The goal of this section is to introduce variational splines on manifolds and to show that such splines can be used for reconstruction of  bandlimited functions from appropriate sets of samples.

Given a $\rho$ lattice $M_{\rho}=\{x_{\gamma}\}$ and a sequence $\{z_{\gamma}\}\in l_{2}$ we
will be
 interested in finding a
 function $s_{k}$ in the Sobolev space $ H^{2k}(M),$ where $k >d/2,\>\>\>d=dim\>M,$  such that
\begin{enumerate}
\item $ s_{k}(x_{\gamma})=z_{\gamma}, x_{\gamma}\in \ M_{\rho};$

\item function $s_{k}$ minimizes functional $g\rightarrow \|L^{k}g\|_{L_{2}(M)}$.
\end{enumerate}

For a given sequence  \  $\{z_{\gamma} \}\in l_{2}$ consider a function $f$
from $H^{2k}(M)$ such that $f(x_{\gamma})=z_{\gamma}.$ Let $\mathcal{P}f$
 denote the orthogonal projection of this function $f$  in the Hilbert
space $H^{2k}(M)$ with the  inner product
$$\left<f,g\right>=\sum_{x_{\gamma}\in
M_{\rho}}f(x_{\gamma})g(x_{\gamma})+ \left<L^{k/2}f, L^{k/2}g\right>$$
on the subspace
$U^{2k}(M_{\rho})=\left \{f\in H^{2k}(M)|f(x_{\gamma})=0\right \}$ with the norm generated 
by the same inner product.
Then the function $g=f-\mathcal{P}f$ will be the unique solution of the
above minimization problem for the
 functional $g\rightarrow \|L^{k}g\|_{L_{2}(M)},
 k=2^{l}d$.

  It is convenient to introduce the so-called Lagrangian splines. For a point $x_{\gamma}$ in a lattice $M_{\rho}$ the corresponding Lagrangian spline $l^{2k}_{\gamma }$ is a function  in $ H^{2k}(M)$ that minimizes the
same functional and
 takes value $1$ at the point $x_{\gamma}$
and $0$ at all other points of $M_{\rho}$.
 Different parts of the following theorem can be found in \cite{Pes00}, \cite{Pes04a}, \cite{Pes0}.

\begin{theorem} The following statements hold:

\begin{enumerate}

  \item for any function $f$ from $H^{2k}(M), \>\>\>k=2^{l}d, \>\>l=1,2,
..., $ there exists a unique
function $s_{k}(f)$ from the Sobolev space $H^{2k}( M), $ such that
$ f|_{M_{\rho}}=s_{k}(f)|_{M_{\rho}}; $ and this function 
  $s_{k}(f)$ minimizes the functional $u\rightarrow \|L^{k}u\|_{L_{2}( M)}$;

\item  every such function $s_{k}(f)$ is of the form 
$$
s_{k}(f)=\sum_{x_{\gamma}\in M_{\rho}}
f(x_{\gamma})l^{2k}_{\gamma};
$$

\item   functions $l^{2k}_{\gamma}$
form a Riesz basis in the space
 of all polyharmonic functions with singularities on $M_{\rho}$ i.e.  in the
space of such functions from
 $H^{2k}( M )$ which in the sense of distributions satisfy equation

$$L ^{2k}u=\sum_{x_{\gamma }\in M_{\rho}}\alpha _{\gamma }\delta (x_{\gamma
})$$ where $\delta (x_{\gamma})$ is the Dirac measure at the point
$x_{\gamma }$; 

\item   if in addition the
set $M_{\rho}$ is invariant under some subgroup of diffeomorphisms acting on
$M$ then every two functions $l^{2k}_{\gamma}, l^{2k}_{\mu}$
 are translates of each other.
 \end{enumerate}
 \label{Splines}
 \end{theorem}

Next, if $f\in H^{2k}( M), k=2^{l}d, l=0,1,...$ then
 $f-s_{k}(f)\in U^{2k}(M_{\rho})$ and we have for
$k=2^{l}
d, l=0,1,...$
  $$\|f-s_{k}(f)\|_{L_{2}( M)}\leq
 (C_{0}\rho)^{k}\|L^{k/2}(f-s_{k}(f))\|_{L_{2}(M)}.$$
Using minimization property of $s_{k}(f)$ we obtain
the inequality 
\begin{equation}
\left\|f-\sum_{x_{\gamma}\in M_{\rho}}f(x{_\gamma})l_{x_{\gamma}}\right\|_{L_{2}(M)}\leq (c_{0}\rho)^{k}\|L^{k/2}f\|_{L_{2}( M)}, k=2^{l}d,\ l=0,1,...,\label{SobAppr}
\end{equation}
and for $f\in E_{\omega}(L)$ the Bernstein inequality gives 
for any $f\in E_{\omega}(L)$ and $ k=2^{l}d, \ l=0,1,....$,
\begin{equation}
\left\|f-\sum_{x_{\gamma}\in M_{\rho}} f(x{_\gamma})l_{x_{\gamma}}\right\|_{L_{2}( M)}\leq(c_{0}\rho\sqrt{\omega} )^{k}\|f\|_{L_{2}( M)}.\label{PWAppr}
\end{equation}

These inequalities lead to the following Approximation  and Reconstruction Theorem.

\begin{theorem}There exist constants $C=C(M)>0$ and  $c_{0}=c_{0}(M)>0$  such that for any $\omega>0$ and any $M_{\rho}$ with 
   $0<\rho \leq c_{0}\omega^{-1/2}$ the following inequality holds for all $f\in E_{\omega}(L)$
$$
\sup_{x\in M}|(s_{k}(f)(x)-f(x))|\leq \omega^{d}
\left(C(M)\rho^{2}\omega\right)^{k-d}\|
f\|, \>k=(2^{l}+1)d,\> l=0, 1, ... .
$$
In other words, by choosing  $\rho>0$ such that
$$
\rho<\left(C(M)\omega   \right)^{-1/2},
$$ 
one obtains  the following reconstruction algorithm  
$$
f(x)=\lim_{k\rightarrow \infty} s_{k}(f)(x),
$$
where convergence holds in the uniform norm.
\end{theorem}

It should be noted that there exists an algorithm \cite{Pes0} which allows to express variational splines in terms of eigenfunctions of the operator $L$. Moreover, it was also shown \cite{Pes04a} that eigenfunctions of $L$ that belong to a fixed space $E_{\omega}(L)$  can be perfectly approximated by  eigenfunctions of certain finite-dimensional matrices in  spaces of splines with a fixed set of nodes.

					\section{Conclusion}

The analysis of functions defined on manifolds is of central importance not only to neuroscience (studies of vision, speech, and motor control), but also to population genetics, finding patterns in gene data \cite{MF}, \cite{MPC}, and manifold models for general signals and images \cite{P}. The present study expands the well-developed field of time-frequency analysis based on wavelets, frames and splines from Euclidean spaces to compact Riemannian manifolds thus giving the means for modeling various important complex phenomena.


\begin{thebibliography}{99}

\bibitem{BKMP1}
P. Baldi,  G. Kerkyacharian, D. Marinucci,  D. Picard, {\em Subsampling needlet coefficients on the sphere},  Bernoulli 15 (2009), no. 2, 438-463. 

\bibitem{BKMP2}
P. Baldi,  G. Kerkyacharian, D. Marinucci,  D. Picard, {\em Asymptotics for spherical needlets},  Ann. Statist. 37 (2009), no. 3, 1150-1171.

\bibitem{SS}
S. Bernstein,  S. Ebert, {\em Wavelets on S3 and SO(3)Ñtheir construction, relation to each other and Radon transform of wavelets on SO(3)},  Math. Methods Appl. Sci. 33 (2010), no. 16, 1895-1909.


\bibitem{BSie}
D. Bassett, F. SiebenhŸhner, {\em Multiscale Network Organization in the Human Brain,}, this volume, chap. XX 

\bibitem{CGSM}
M. Calixto, J. Guerrero, J. C. S‡nchez-Monreal, {\em Sampling theorem and discrete Fourier transform on the hyperboloid},  J. Fourier Anal. Appl. 17 (2011), no. 2, 240-264. 

\bibitem{CO1}
J. Christensen, G. Olafsson, {\em  Examples of coorbit spaces for dual pairs}, Acta Appl. Math. 107 (2009), no. 1-3, 25-48



\bibitem{CO2}
J. Christensen, G. Olafsson, {\em Coorbit spaces for dual pairs}, Appl. Comput. Harmon. Anal. 31 (2011), no. 2, 303-324.

\bibitem{C}
J. Christensen, {\em Sampling in reproducing kernel Banach spaces on Lie groups},  J. Approx. Theory 164 (2012), no. 1, 179-203. 


\bibitem{CKP}
T. Coulhon, G. Kerkyacharian, P. Petrushev, {\em 
    Heat kernel generated frames in the setting of Dirichlet spaces},  arXiv:1206.0463.

\bibitem{D}
I. Daubechies, {\em Ten Lectures on Wavelets},  Pennsylvania, Philadelphia (1992).



\bibitem{FM1}
F. Filbir, H. Mhaskar,  {\em A quadrature formula for diffusion polynomials corresponding to a generalized heat kernel},  J. Fourier Anal. Appl.  16  (2010),  no. 5, 629-657.


\bibitem{FuhrM}
H. F\"{u}hr, A.Mayeli, {\em 
Homogeneous Besov spaces on stratified Lie groups and their wavelet characterization}, will appear in J. Function spaces and Applications, 2012.

\bibitem{FuhrG}
H. F\"{u}hr,  K. Grochenig,  {\em Sampling theorems on locally compact groups from oscillation estimates},  Math. Z. 255 (2007), no. 1, 177-194.

\bibitem{HF}
H. F\"{u}hr, {\em Abstract harmonic analysis of continuous wavelet transforms},  Lecture Notes in Mathematics, 1863. Springer-Verlag, Berlin, 2005. x+193 pp. ISBN: 3-540-24259-7 

\bibitem{FY} 
M. Frazier and B. Jawerth, {\em Decomposition of Besov Spaces}, Ind. Univ. Math. J. 
{\bf 34} (1985), 777-799.

\bibitem{gm1} D. Geller and A. Mayeli, {\em Continuous Wavelets on Compact Manifolds},  Math. Z. {\bf 262} (2009), 895-92.

\bibitem{gm3} D. Geller and A. Mayeli, {\em Nearly Tight Frames and Space-Frequency Analysis on Compact Manifolds} (2009), 
Math. Z. {\bf 263} (2009), 235-264.

\bibitem{gm4} D. Geller and A. Mayeli, {\em Besov spaces and frames on compact manifolds}, Indiana Univ. Math. J. 58 (2009), no. 5, 2003-2042.


\bibitem{GM100} D. Geller and D. Marinucci, {\em Mixed needlets}, J. Math. Anal. Appl. 375 (2011), no. 2, 610-630.

\bibitem{GM2} D. Geller,  and D. Marinucci, (2010) Spin Wavelets on the Sphere, J. of Fourier Analysis and its Applications, Vol. 6, pp.840-884, arxiv: 0811.2935.

\bibitem{gpes} D. Geller and I. Pesenson, {\em Band-limited localized Parseval frames and Besov spaces on compact homogeneous manifolds}, J. Geom. Anal. 21 (2011), no. 2, 334-37.

\bibitem{gm5}
D. Geller, A. Mayeli,  {\em  Wavelets on manifolds and statistical applications to cosmology}, Wavelets and multiscale analysis, 259-277, Appl. Numer. Harmon. Anal., BirkhŠuser/Springer, New York, 2011. 


\bibitem {H2}
S. Helgason, {\em Groups and Geometric Analysis}, Pure and Applied Mathematics, 113. Academic Press, Inc., Orlando, FL, 1984. xix+654 pp. ISBN: 0-12-338301-3.


\bibitem{Hor} L. H\"{o}rmander, {\em The analysis of linear partial differential operators. III. Pseudo-differential operators},   Springer, Berlin, 2007. viii+525 pp. ISBN: 978-3-540-49937-4.

\bibitem{MF}
Y. Ma and Y. Fu, {\em Manifold learning theory and applications}, CRC, London 2012.


\bibitem{M1}
S. Mallat, {\em A Wavelet Tour of Signal Processing, the sparse way}, Academic
Press, 3rd edition, 2008.

\bibitem{M2}
S. Mallat, {\em Group Invariant Scattering}, to appear in
ÒCommunications in Pure and Applied MathematicsÓ, 2012,
http://arxiv.org/abs/1101.2286.

\bibitem{MP}
D. Marinucci,  G. Peccati, {\em  Random fields on the sphere. Representation, limit theorems and cosmological applications}, London Mathematical Society Lecture Note Series, 389. Cambridge University Press, Cambridge, 2011. xii+341 pp. ISBN: 978-0-521-17561-6.

\bibitem{M-all}
D.Marinucci et al., {\em Spherical Needlets for CMB Data Analysis}, Monthly Notices of the Royal Astronomical Society, Vol. 383, (2008),  pp. 539-545.
 
\bibitem{A}
A. Mayeli, {\em  Shannon multiresolution analysis on the Heisenberg group}, J. Math. Anal. Appl. 348 (2008), no. 2, 671-684.

\bibitem{AI}
A. Mayeli, I. Pesenson, {\em Space-Frequency localized wavelets for spherical Besov spaces on 
the Heisenberg group}, submitted. 


\bibitem{MPC}
P. Menozzi, A. Piazza, L. Cavalli-Sforza , {\em Synthetic Maps of Human Gene Frequencies in Europeans}, SCIENCE, vol. 201, 1, 1978.


\bibitem{Mum}
D. Mumford, {\em Neuronal architectures for pattern-theoretic problems},  In
Large-scale neuronal theories of the brain. Edited by Koch, C. and Davis, J. L.
Cambridge, MA: MIT Press,  (1994), 125Ð152.


\bibitem{NPW} F.J. Narcowich, P. Petrushev and J. Ward, {\em Localized Tight frames on spheres},
SIAM J. Math. Anal. 38, (2006), 574-594.


\bibitem{OP}
J. Ortega-Cerd\'{a}, B. Pridhnani, {\em Beurling-Landau's density on compact manifolds }, Journal of Functional Analysis, Volume 263, Issue 7,  (2012), 1825-186.

\bibitem{NSI}
P. Ninez, R. Srinivasan, L. Ingber, {\em Theoretical and experimental electrophysiology in human neocortex}, this volume, chap. XX.


\bibitem{Pes98}
I. Pesenson, {\em Sampling of Paley-Wiener functions on stratified
groups}, J. of Fourier Analysis and Applications {\bf 4} (1998),
269--280.


\bibitem{Pes00}
I. Pesenson, {\em A sampling theorem on homogeneous manifolds},
Trans. Amer. Math. Soc. {\bf 352} (2000), no. 9, 4257--4269.

\bibitem{Pes04a}
I.~Pesenson, {\em An approach to spectral problems on Riemannian
manifolds,}  Pacific J. of Math. Vol. 215(1), (2004), 183-199.

\bibitem{Pes04b}
I.~Pesenson,  {\em Poincare-type inequalities and reconstruction
of Paley-Wiener functions on manifolds, }  J. of Geometric Analysis
, (4), 1, (2004), 101-121.


\bibitem{Pes0}
I. Pesenson, {\em Variational splines on Riemannian manifolds
with applications to integral geometry}, Adv. in Appl. Math. 33
(2004), no. 3, 548--572.


\bibitem{Pes6}
I. Pesenson, {\em Deconvolution of band limited functions on
symmetric spaces},  Houston J. of Math., 32, No. 1, (2006),
183-204.



\bibitem{Pes07}
I. ~Pesenson, {\em Frames in Paley-Wiener spaces on Riemannian
manifolds}, in Integral Geometry and Tomography, Contemp. Math.,
405, AMS, (2006), 137-153.


\bibitem{Pes09a}
I. ~Pesenson, {\em A Discrete Helgason-Fourier Transform for
Sobolev and Besov functions on noncompact symmetric spaces},
Contemp. Math, 464, AMS, (2008), 231-249.


\bibitem{Pes09}
 I. ~Pesenson, {\em Paley-Wiener approximations and multiscale approximations in
Sobolev and Besov spaces on manifolds,}  J. of Geometric Analysis
, 4, (1), (2009), 101-121.


\bibitem{Pes11}
I. Pesenson, M. Pesenson, {\em Approximation of Besov vectors by Paley-Wiener vectors in Hilbert spaces},
  Approximation Theory XIII: San Antonio 2010 (Springer Proceedings in Mathematics), by Marian Neamtu and Larry Schumaker, 249--263.


\bibitem{Pessubm}
 I. ~Pesenson, {\em Localized Bandlimited nearly tight frames and Besov spaces on domains in Euclidean spaces}, submitted,  	arXiv:1208.5165v1.
 
 
\bibitem{Pesssubm}
 I. ~Pesenson, {\em Frames  and Besov spaces on non-compact manifolds}, submitted.

\bibitem{PM}
M. Pesenson, {\em Adaptive Multiscale Encoding - a Computational Function of Neuronal Synchronization}, this volume, chap. XX.

\bibitem{PX}
P. Petrushev,  Y. Xu, {\em Localized polynomial frames on the ball}, Constr. Approx. 27 (2008), no. 2, 121-148.

\bibitem{P}
G. PeyrŽ, {\em Manifold models for signals and images}, Computer Vision and Image Understanding, 113 (2009) 249-260.

\bibitem{Tay81} M. Taylor, {\em Pseudodifferential Operators}, Princeton University
Press, 1981.

\bibitem{TSL}
J. Tenenbaum, V. de Silva, J. Langford, {\em A Global Geometric Framework for Nonlinear Dimensionality Reduction}, SCIENCE vol. 290, 22, 2000.


\end{thebibliography}
\end{document}